\documentclass[article]{aa520} % for a referee version
\usepackage{graphicx}
\begin{document}
   \title{Mixing of shear Alfv\'en wave packets }

   \author{Nicolas Bian
          \inst{1}
          \and David Tsiklauri
          \inst{2}}

   \institute{Jodrell Bank Centre for Astrophysics, School of Physics and Astronomy, University of Manchester,
   Manchester, M60 1QD, UK
              \email{nbian@hotmail.com}
         \and
            Institute for Material Research, University of Salford, Greater Manchester, M5 4WT, UK }

   \date{Received ; accepted}

% \abstract{}{}{}{}{}
% 5 {} token are mandatory

  \abstract{The propagation of shear Alfv\'en wave packets in inhomogeneous magnetic
  fields, at the origin of their distortion, regardless of the occurrence of non-linear coupling,
  is considered.
  It is shown that the distortion mechanism can be regarded as mixing process and hence, standard
"phase mixing" corresponds to the effect of an "Alfv\'enic"
shear flow while enhanced dissipation at a magnetic X-point
corresponds to mixing by an "Alfv\'enic" strain flow. 
The evolution of the global wave field
is supposed to
result from the dynamics of a superposition of wave packets and a kinetic equation for the wave energy is obtained
following this eikonal (WKB) description.
Since shear Alfv\'en wave packets experience continuous
shearing/straining while transported by an inhomogeneous Alfv\'enic
flow $\mathbf{V}_{A}$, their mixing process, in physical space, is also a
cascade of wave energy in k-space. The wave energy spectrum resulting from
this linear mechanism of energy transfer is determined
for the special case of waves propagating
along chaotic magnetic field lines, the analog of a chaotic mixing process.
 The latter follows a
$k^{-1}$ power-law, in the energy conserving range in $k$ space.
}

\titlerunning{Mixing of shear Alfv\'en wave packets}
\authorrunning{Bian and Tsiklauri}

   \maketitle
%
%________________________________________________________________

\section{Introduction}
Phase mixing, the propagation and enhanced dissipation of shear
Alfv\'en waves, remain active areas of research in studies of the
solar corona, both in relation with its heating mechanism and the
acceleration of the fast solar wind; see e.g. reviews on the
subjects (Browning 1989; Walsh and Ireland 2003; Ofman 2005) and references within.
The reason is that if Alfv\'en waves are excited and dissipate some of their energy in the
solar corona, this contributes to the overall coronal energy budget.

Due to the very low collisionality of the medium, a main difficulty of waves
dissipation theories is, however, the justification of how these
fluctuations can thermalize, i.e. reach the very small dissipative scale,
before they leave the corona. But since the coronal
plasma is highly inhomogeneous, it is quite natural to invoke waves interaction with the ambient
background as a reason for the creation of fine enough scales in the wave field.
It is in this context that,
phase mixing of Alfv\'en waves was proposed, by Heyvaerts and Priest
(1983), as an enhanced dissipation mechanism. They considered simple
inhomogeneity of the medium, in one direction
perpendicular to the unidirectional background magnetic field. The
existence of a shear in the Alfv\'en velocity $\mathbf{V}_{A}$ results
in the creation of progressively smaller transverse scales in the
wave field while it propagates and is transported along the unperturbed magnetic
field. In this case, the build-up of a diffusive scale is linear
in time ($dk/dt\propto cst$), resulting in a dissipation time proportional to
$S^{1/3}<<S$. Here, $S$ is the large Lundquist number $S\equiv
\tau_{\eta}/\tau_{A}\gg 1$ of the plasma. The corresponding
diffusive scale is $S^{-1/3}\ll1$. Therefore, the dissipation
time is found to me much shorter than the very large diffusion time
$\tau_{\eta}$, proportional to $S\gg1$, for the wave to dissipate
in the absence of inhomogeneity.

More recent studies have shown that the creation of a dissipative
scale can be made exponentially fast ($dk/dt\propto cst.k$), with associated time of
the order of $\ln S << S^{1/3} << S$ (Similon and Sudan 1989;
Petkaki et al. 1998; Malara et. al 2007). The latter improved
efficiency of the dissipation results essentially from the
consideration of a more complex magnetic geometry, with the property
of local exponential separation or divergence of neighboring
magnetic field lines. In this case, the diffusive length scale is
proportional to $S^{-1/2}$. Therefore, fast dissipation is possible
in three-dimensional chaotic magnetic fields (Similon and Sudan
1989, Petkaki et al 1998, Malara et. al 2000, Malara et. al 2007)
and in two-dimensional regular ones (Smith et al. 2007, Malara et
al. 2003, Ruderman et. al 1998), provided they possess a certain
degree of complexity.

One main goal of this work is to show that the propagation and distortion
of shear Alfv\'en wave packets, in
inhomogeneous magnetic fields, can be regarded as a "mixing" effect.
As is well known, the kinematics of mixing are governed by
the topological properties of the flow. It naturally follows that
the distortion of Alfv\'en wave packets,
propagating in complex magnetic fields, is the counterpart of a
chaotic mixing process(Ottino 1988).

The interaction among shear Alfv\'en waves
is also a well-studied mechanism of energy transfer to small scales, the non-linear distortion
resulting from elastic collisions between counter-propagating wave packets
producing a flux of their energy to small dissipative scales.
Yet, linear distortion through interaction with the
background produces a similar flux.
In the following, we neglect non-linearities
in the magnetohydrodynamics equations. The reason is to focus primarily on the
spectral properties of this energy transfer for waves propagating
in complex magnetic fields.

We obtain a kinetic equation for the wave energy
from a representation of the global
wave perturbation as an ensemble wave packets (Similon and Sudan 1989). Hence, the
wave energy spectrum F(k,t) is the sum of the energy of each
individual wave packet, with associated wave vector and position,
whose time dependence is given by Hamiltonian ray equations. This
WKB method has already been employed by several authors (Petkaki et al 1998,
Malara et. al 2000, Malara et. al 2007) to study shear Alfv\'en waves
propagation and dissipation in different types of coronal magnetic structures. Our
study is a continuation of these works. In section 4, we determine
the transient and stationary energy spectra, F(k,t) and F(k), for the special
case of non-interacting Alfv\'en waves propagating in a synthetic chaotic
magnetic configuration with quasi-uniform rate of
field lines exponentiation (Similon and Sudan 1989). In section 3, we show
the correspondence which exists between the problem of the enhanced
dissipation of Alfv\'en waves, transported by inhomogeneous Alfv\'enic
flows, and the problem of mixing (Ottino 1988), reviewed in section 2. Conclusions are
given in the last section of this article.

\section{Mixing of a passive tracer}

The mixing of a passive field $\psi$ advected by a
flow $\mathbf{U}$ is described by the linear advection-diffusion
equation
\begin{equation}\label{eq1}
\frac{\partial \psi}{\partial t}+\mathbf{U}.\nabla \psi=\eta
\nabla^{2}\psi
\end{equation}
In absence of flow, $\mathbf{U}=0$, the field $\psi$ only diffuses
on a large time scale, for small diffusion coefficient $\eta$,
because this time is proportional to $\eta^{-1}$. If the flow has
some spatial variation, i.e. it is inhomogeneous, the advected field
is stirred. It then develops increasing gradients, until diffusion
takes over and ultimately mixes the field. Because diffusion alone
is slow to mix, a stirring flow is required to accelerate the
dissipative process. An important issue is therefore
to find flows $\mathbf{U}$ that have the property of being good
mixers. For the enhanced dissipation of shear-Alfv\'en waves, this
property concerns the Alfv\'enic flow $\mathbf{V}_{A}$, which by
definition, is responsible for the advection of the wave field. To
show this, we first concentrate on two standard and simple flow
configurations having their counterpart in the problem of
dissipation of Alfv\'en waves, and compare their mixing efficiency.

Consider the effect of shear flow $\mathbf{U}=(U_{0}'y,0,0)$,
\begin{equation}\label{eq2}
\frac{\partial \psi}{\partial t}+U'_{0}y\frac{\partial
\psi}{\partial x}=\eta(\frac{\partial^{2} \psi}{\partial
^{2}x}+\frac{\partial^{2} \psi}{\partial ^{2}y}+\frac{\partial^{2}
\psi}{\partial^{2}z})
\end{equation}
with initial condition $\psi(x,y,t=0)=\psi_{0}\sin k_{0}x$, which
has no dependence in $y$. When $\eta=0$, the field is just advected,
resulting in $\psi(x,y,t)=\sin k_{0}(x-U_{0}'yt)$, i.e an unbounded
creation of smaller scales along $y$, the shear-wise direction. When
$\eta \neq 0$, a dissipative cut-off is introduced. In this case, we
can look for a solution of the form $\psi=\psi_{0}f(t)\sin
k_{0}(x-U'_{0}yt)$, which gives
\begin{equation}
f(t)=\exp(-\eta k_{0}^{2}(t+\frac{U_{0}'}{3}t^{3})),
\end{equation}
hence, the initial field evolves according to
\begin{equation}
\psi(x,y,t)=\psi_{0}\sin k_{0}(x-U'_{0}t)\exp (-\eta
k_{0}^{2}(t+\frac{U_{0}'^{2}}{3}t^{3}))
\end{equation}
From the above solution, we obtain that, for $\eta\ll 1$, the mixing
time scale associated with the shear flow is
\begin{equation}
\tau_{shear} \propto (\eta U_{0}'^{2}k_{0}^{2})^{-1/3}
\end{equation}
The associated dissipative length $l_{d}$ is found from
Eq.(\ref{eq2}), by balancing the advection term, the second on its
left, with the diffusion (along $y$) term, the second on its right,
$U_{0}'l_{d}(\psi/L)\sim \eta (\psi/l_{d}^{2})$:
\begin{equation}
l_{d}\sim (\frac{\eta L}{U_{0}'})^{1/3}
\end{equation}
By performing a formal spectral transformation of Eq.(\ref{eq2}) it
is also found that individual modes evolve according to
\begin{equation}
\frac{\partial \widehat{\psi}_{k}}{\partial
t}+\mathbf{U}_{0}'k_{y}\frac{\partial \widehat{\psi}_{k}}{\partial
k_{x}}=-\eta (k_{x}^{2}+k_{y}^{2})\widehat{\psi}_{k},
\end{equation}
with the right-hand side of the above expression representing
advection in $k$-space. This also indicates that a shear flow
results in a spectral expulsion towards large absolute value
shear-wise wave numbers, corresponding to an increase tilting of the
phase front in the flow-wise direction. Hence, the physics behind
the accelerated diffusion of the tracer $\psi$ in a shear flow is
intuitively analogous to the one of \emph{phase mixing}, which
occurs for $\mathbf{V}_{A}=(V_{A}'y,0,0)$. To gain a further insight
into the problem, let us transform the advection-diffusion equation,
by introducing the "shearing coordinates" that follows the
streamlines :
\begin{equation}
[x'=x-U_{0}'yt,y'=y,t'=t]
\end{equation}
From the chain rule for derivatives :
\begin{equation}
[\frac{\partial}{\partial x}=\frac{\partial}{\partial
x'},\frac{\partial}{\partial y}=-U_{0}'t\frac{\partial}{\partial
x}+\frac{\partial}{\partial y'},\frac{\partial}{\partial
t}=-U_{0}'y\frac{\partial}{\partial x}+\frac{\partial}{\partial
t'}],
\end{equation}
the advection-diffusion is now described by
\begin{equation}
\frac{\partial \psi}{\partial t}=\eta(\frac{\partial^{2}}{\partial
^{2}x}+(-U_{0}'t\frac{\partial}{\partial x}+\frac{\partial}{\partial
y'})^{2})\psi
\end{equation}
Keeping only the dominant terms in this equation, leads, after
introduction of the new variable $\tau=(U_{0}'^{2}t^{3})/3$, to a
diffusion equation :
\begin{equation}
\frac{\partial \psi}{\partial \tau}=\eta
\frac{\partial^{2}\psi}{\partial^{2} x}
\end{equation}
As is well known, there are two self-preserving solutions of a
diffusion equation: harmonic and gaussian. Their temporal evolution
is different, however, their damping time is identical. Indeed,
contrary to the harmonic solution studied above, an initial gaussian
packet has its amplitude that decays as $1/(\eta\tau)^{1/2}$, while
it spreads. For the phase mixing problem, such power law behavior of
localized initial wave-packets, i.e its diffusion along field lines, was noticed and studied by Hood et
al. (2002); see also Tsiklauri et al. (2003).

Now let us consider passive advection in a simple strain
$\mathbf{U}=(-\alpha x,\alpha y, 0)$ with harmonic initial condition
independent of $z$:
\begin{equation}\label{eq13}
\frac{\partial \psi}{\partial t}- \alpha x\frac{\partial
\psi}{\partial x}+ \alpha y\frac{\partial \psi}{\partial
y}=\eta(\frac{\partial^{2} \psi}{\partial ^{2}x}+\frac{\partial^{2}
\psi}{\partial ^{2}y}+\frac{\partial^{2} \psi}{\partial ^{2}z})
\end{equation}
A solution of the form
\begin{equation}
\psi=\psi_{0}f(t)\sin (k_{x}(t)x+k_{y}(t)y)
\end{equation}
is sought for, and hence,
\begin{eqnarray}
&&\dot{k_{x}}=k_{x}\alpha; \:\:\:\ \dot{k_{y}}=-k_{y}\alpha,
\\&&k_{x}(t)=k_{0}\exp \alpha t; \:\:\:\ k_{y}(t)=k_{0}\exp -\alpha t,
\end{eqnarray}
\begin{eqnarray}
&&\dot{f}=-\eta(k_{x}^{2}+k_{y}^{2})f \\&&f(t)=\exp[-\frac{\eta
k_{0}^{2}}{2\alpha}(\exp^{2\alpha t} - 1))]
\end{eqnarray}
The field amplitude has super-exponential decay in time. It follows
that for $\alpha>0$, the mixing time due to a strain $\tau_{strain}$
is reduced compared to $\tau_{shear}$, i.e.
\begin{equation}
\tau_{strain}\propto (1/\alpha) \ln(\alpha/\eta k_{0}^{2}),
\end{equation}
the reason is, that for a strain, the wave vector increases
\emph{exponentially } with time and not linearly. And we can
anticipate that this situation corresponds to enhanced dissipation
of Alfv\'en waves in the presence of a magnetic X-point configuration
with $\mathbf{V}_{A}=(\alpha x, \alpha y,0)$. This result remains
unchanged for the three dimensional stagnation flow
$\mathbf{U}=(\alpha x,\alpha y,-2\alpha z)$. The dissipative scale
length for advection and diffusion to balance each other in
Eq.(\ref{eq13}), is given by
\begin{equation}
l_{d}\sim (\frac{\eta}{\alpha})^{1/2}
\end{equation}

The general problem of the mixing of a passive scalar, in arbitrary
flows, can be approached from a WKB description (Antonsen 1996).
This is done, for instance, by introducing a filter,
$\widehat{\psi}(\mathbf{k},\mathbf{x},t)=\int d^{3}x
\exp[-\mathbf{x}^{2}/\lambda^{2}+
i\mathbf{k}(\mathbf{x}-\mathbf{x'})]\psi(\mathbf{x'},t)$,
with support $\lambda \ll L$, $L$ being the length scale of
variation of $\mathbf{U}$. It is a local Fourier transform in
$\mathbf{x}$. Applying this transformation to the
advection-diffusion equation (\ref{eq1}), we obtain
\begin{equation}\label{eq24}
D_{t}\widehat{\psi}=-\eta k^{2}\widehat{\psi}
\end{equation}
with total derivative $D_{t}=\partial
_{t}+\dot{\mathbf{x}}\nabla+\dot{\mathbf{k}}\nabla_{k}$ and
$k^{2}=k_{x}^{2}+k_{y}^{2}+k_{z}^{2}$. Equation (\ref{eq24}), which
describes transport in $(\mathbf{x},\mathbf{k})$ space is also known
as the kinetic wave equation in plasma physics (Sagdeev and Galeev
1969) and includes non-linear coupling terms on its
right-hand side. It is equivalently obtained by use of the Wigner
distribution (Bastiaans 1979, Antonsen 1996). The respective
characteristics are :
\begin{equation}\label{eq25}
\dot{\mathbf{x}}=\mathbf{U}=\nabla_{k}H
\end{equation}
\begin{equation}\label{eq26}
\dot{\mathbf{k}}=-\nabla(\mathbf{k}.\mathbf{U})=-\nabla H
\end{equation}
with $H=\mathbf{k}.\mathbf{U}$. The general solution of the kinetic
wave equation (\ref{eq24}) can be expressed in term of phase space
trajectories $\mathbf{\xi}(\mathbf{x}',t)$ and
$\mathbf{\kappa}(\mathbf{x}',\mathbf{k}',t)$ where
$d\mathbf{\xi}/dt=\mathbf{v}(\xi,t)$ and
$d\mathbf{\kappa}/dt=-\nabla\mathbf{v}(\xi,t).\mathbf{\kappa}$, and
$\mathbf{\xi}(\mathbf{x}',0)=\mathbf{x}',
\mathbf{\kappa}(\mathbf{x}',\mathbf{k}',0)=\mathbf{k}'$. The
solution is $\widehat{\psi}(\mathbf{x},\mathbf{k},t)=\int
d^{3}x'd^{3}k' \widehat{\psi}(\mathbf{x}',\mathbf{k}',0)
\delta(\mathbf{x}-\mathbf{\xi}(\mathbf{x'},t))
\delta(\mathbf{k}-\mathbf{\kappa}(\mathbf{x}',\mathbf{k}',t))\exp(-\eta\int_{0}^{t}\kappa^{2}dt')$.

The physical interpretation of the WKB approach to mixing is clear.
It means that the advected field is decomposed into an ensemble of
"packets" with length scale, $\lambda$, much smaller than the one,
$L$, characterizing the spatial variation of the flow $\mathbf{U}$.
Therefore, each packet experiences, along its trajectory, a velocity
field $\mathbf{U}$ which is a linear function of the space
coordinates :$\mathbf{U}(\mathbf{x}_{p}(t))+
\partial\mathbf{U}/\partial\mathbf{x}\mid_\mathbf{x_{p}(t)}.\mathbf{x}$,
$\mathbf{x}_{p}(t)$ being the position of the packet at time $t$.
An important quantity, therefore, is the local rate of strain tensor,
 $\partial\mathbf{U}/\partial \mathbf{x}$, of the flow. To
sum-up, the packets are deformed only by the shear and the strain
considered below. This brings us to the analogous effect for Alfv\'en
wave packets.

\section{Mixing of shear Alfv\'en waves}

The rays equations of optics are $\dot{
\mathbf{x}}=\nabla_{k}\omega$ and $\dot{ \mathbf{k}}=-\nabla
\omega$, with $\omega(\mathbf{k})$ the wave dispersion relation. The
eikonal description of MHD waves was introduced by Weinberg (1962);
see also Similon and Sudan (1989), Petkaki et al. (1998), who used it in the
context of wave dissipation in the corona. For the non-dispersive
shear Alfv\'en waves, with $\omega=\pm\mathbf{k}.\mathbf{V_{A}}$, the
rays equations read
\begin{equation}\label{eq32}
\dot{ \mathbf{x}}=\pm\mathbf{V}_{A},
\end{equation}
\begin{equation}\label{eq33}
\dot{ \mathbf{k}}=\mp\nabla(\mathbf{k}. \mathbf{V}_{A}).
\end{equation}
with the energy equation being
\begin{equation}\label{eq34}
\dot{e}_{\pm}=-\eta k^{2} e_{\pm},
\end{equation}
The latter is required to study the energetics of the mixing
process. The equation of motion Eq.(\ref{eq32}) states that
shear-Alfv\'en wave packets propagate along magnetic field lines so
they disperse as the field lines do. It provides information on the
precise path along which energy is deposited. The enhanced
dissipation time follows from equation (\ref{eq33}), giving the rate
of stretching of wave packets, i.e. the variation of their
wave-vector, when combined with the energy equation. Dispersion and
stretching are however related, due to the following consideration
of two infinitesimally separated trajectories and the phase
gradient, $\nabla \psi \equiv \mathbf{k}$, they generate in an
inhomogeneous Alfv\'enic flow $\mathbf{V}_{A}$. Indeed, constancy of
the phase along a trajectory implies that
$\psi(\mathbf{\xi}(\mathbf{x}+\mathbf{r},t),t)-\psi(\mathbf{\xi}(\mathbf{x},t),t)=\delta
\mathbf{\xi}(\mathbf{x},t).\nabla
\psi(\mathbf{\xi}(\mathbf{x},t),t)=const$ with $\delta
\mathbf{\xi}(\mathbf{x},t)=\delta
\mathbf{\xi}(\mathbf{x}+\mathbf{r},t)-\delta
\mathbf{\xi}(\mathbf{x},t)$ being the differential separation
vector. Thus, $ d(\delta\mathbf{\xi}.\mathbf{k})/dt=0$. According to
Eq.(\ref{eq32}), the trajectory of a wave packet is
$\dot{\mathbf{\xi}}=\pm\mathbf{V}_{A}(\mathbf{\xi},t)$ and therefore
$d\delta\mathbf{\xi}(\mathbf{x},t)/dt=\pm\delta
\mathbf{\xi}.\nabla\mathbf{V}_{A}(\mathbf{\xi}(\mathbf{x},t),t)$.
Hence, the stretching properties of the Alfv\'enic flow, represented
by Eq.(\ref{eq33}) for the wave vector are related to the dispersion
of magnetic field lines. Considering the phase mixing situation,
with an Alfv\'enic shear flow $\mathbf{V}_{A}=V'_{A}y \mathbf{x}$, $
dk_{y}/dt=k_{x}V'_{A}$ with solution $k_{y}(t)=k_{x}V_{A}'t$ and
integrating the energy equation, the typical time scale for the
phase mixing is found to correspond indeed to $\tau_{shear}$, i.e
\begin{equation}
\tau_{mix}\propto S^{1/3}\ll S
\end{equation}
with $S$ the Lundquist number $S\gg1$. In the same way, for the
enhanced dissipation of shear Alfv\'en waves at a magnetic X-point,
which corresponds to a an Alfv\'enic strain flow
$\mathbf{V}_{A}=\alpha x\mathbf{x} -\alpha y\mathbf{y}$, $\alpha
\equiv V_{A}'$, the component of the wave vector in the direction of
contraction $\mathbf{x}$, increases now exponentially with time,
$\mathbf{k}_{x}(t)=k_{0}\exp(\alpha t)$. Integrating the energy
equation results in the above mentioned superexponential decay and,
therefore, the enhanced dissipation time corresponds to
$\tau_{strain}$,
\begin{equation}
\tau_{mix}\propto \ln S\ll S^{1/3}
\end{equation}
This dependance of the dissipation time with Lunquist number, for
Alfv\'enic strain, is also typical of wave damping in complicated
magnetic geometries, either if the magnetic field is stochastic
(Similon and Sudan 1989) or regular, but possesses separators
(Malara et al. 2003; Malara et al. 2007). The physics is however the
same, and is related to the divergence of nearby magnetic field
lines. In all these cases, the dissipation time of a wave packet
weakly depends on the Lunquist number, only as a logarithm,
$\tau_{mix}\propto \ln S$, and not as a power law,
$\tau_{mix}\propto S^{1/3}$ obtained for unidirectional magnetic
field, i.e. standard phase mixing. It is now clear that the
propagation of shear Alfv\'en waves in an inhomogeneous magnetic field
is equivalent to a mixing process. Since the kinematics of mixing
are governed by the topological properties of the flow, it results
that the enhanced dissipation of Alfv\'en waves propagating in a
chaotic magnetic field is equivalent, from the energetics viewpoint,
to a chaotic mixing process. The spectral property of the wave energy
flux, in this linear process, is the subject of the following
section.
\section{wave energy spectrum in chaotic magnetic fields}

Since within the eikonal formulation the global wave perturbation
can be treated as a superposition of wave packets with wave-vector
$\mathbf{k}_{j}$, each carrying an energy $e_{j}(t)$, we can also
write the wave energy spectrum in the following form:
\begin{equation}\label{eq40}
F(k,t)=\sum F_{j}(k,t)=\sum e_{j}(t)\delta(k-|\mathbf{k}_{j}(t)|)
\end{equation}
In the presence of a source of wave energy,
continuous shearing/straining experienced by wave packets along
their trajectory leads to a cascade of their energy, and hence, of the
energy of the global wave perturbation. This cascade can be studied
in two different ways. In the first, there is no steady source, but
only an initial distribution of wave packets and energy in space. For this initial
value problem, the interest is in the subsequent transient evolution
of the wave energy spectrum $F_{I}(k,t)$ with initial condition
$F_{I}(k,t=0)=\sum e_{j}(t=0)\delta(k-|\mathbf{k}_{j}(t=0)|)$. As
time increases, each wave number $k_{j}$ increases in time (on average), evolves
toward the dissipation range and, therefore, the wave energy decays.
A second approach consists in taking into account continuous
injection of wave energy given by the ongoing
perturbations. Therefore, wave packets and energy are supposed to be
continually injected at intermediate $k$ values, around $k_{0}$, a
typical injection scale, with $L^{-1}\ll k_{0}\ll k_{d}$ and with
$k_{d}$ the very small dissipative cutoff and $L$ the global length
scale associated with the magnetic field. In this case,
electromechanical energy injection, say the photospheric
drive, becomes balanced by heating and dissipation at high $k$, and a
time-averaged stationary spectrum $F_{S}(k)$ builds-up. We 
turn to the study of the wave energy spectrum that results from
quasi-uniform Alfv\'enic strain experienced by wave-packets as they
propagate in a field with destroyed magnetic surfaces, i.e. along
chaotic magnetic field lines. This configuration, which is a
paradigm in transport theory, for the calculation of anomalous
diffusion coefficients (Rechester and Rosenbluth 1977), was first
considered in the context of coronal wave dissipation by Similon and
Sudan (1989), although they were only interested in the mixing time
scale. Calling $\varepsilon$ the input wave energy flux, dimensional
analysis yields
\begin{equation}
\varepsilon\propto V_{A}'k F_{S}(k),
\end{equation}
with $F_{S}(k)$ the stationary wave energy spectrum. Hence,
\begin{equation}\label{eq42}
F_{S}(k)\propto k^{-1}V_{A}'^{-1}\varepsilon,
\end{equation}
in the  range where dissipation is negligible, i.e. for
$k<k_{d}=(V_{A}'/\eta)^{1/2}$. This relation (\ref{eq42}) for the
wave-energy, which is again derived below from a kinetic equation, is the same as the
Batchelor's spectrum (Batchelor 1959) for scalar variance, obtained
for a flow with uniformly distributed rate of strain, i.e. with
chaotic streamlines.

An other derivation of (\ref{eq42}) follows from writing down the
evolution equation for the wave energy spectrum $F(k,t)$. Taking the
time derivative of Eq.(\ref{eq40}), we obtain
\begin{equation}\label{eq44}
\frac{\partial F(k,t)}{\partial t}=-\eta k^{2}F(k,t)+
\frac{\partial}{\partial k}[V_{A}'k F(k,t)],
\end{equation}
The first term on the right-hand side represents the effect of
dissipation, $\dot{e}=-\eta k^{2}e$, and vanishes for
$\eta=0$. The second term describes transport in k-space for wave
packets that experience uniform rate of strain, $V_{A}'=const$, i.e.
$\dot{k}(t)=V_{A}'k$. The stationary solution of
(\ref{eq44}) for $\eta=0$ is again $F_{s}(k)\propto k^{-1}$. For
$\eta\neq0$, the stationary solution to Eq.(\ref{eq44}), with energy
source $\epsilon$ added at wave numbers where wave energy is fed
into the system, is
\begin{equation}
F_{S}(k)=\frac{\epsilon}{V_{A}'k}\exp(-\eta k^{2}/\alpha)
\end{equation}
This shows the existence of a gaussian dissipative fall-off, for
$F_{s}(k)$, at large $k>k_{d}$.

Let us note that while the derivation of the kinetic
equation (\ref{eq44}) is based on the WKB approximation, the dimensional
analysis obtention of the $k^{-1}$ stationary spectrum does not rely on the assumption of a separation of scales.
The latter is based only on the fact that the wave
energy transfer time is a constant, independent of the scale, given by $V_{A}'$, the quasi-uniform rate of strain
experience by the waves.

Finally, we go back to the initial
value problem, for transient forcing. Discarding dissipation in Eq.(\ref{eq44}), and
changing to the new variables, $\xi=\ln(k)$ and $G=kF(k)$, we can
write Eq.(\ref{eq44}) as
\begin{equation}
\frac{\partial G}{\partial t}=V_{A}'\frac{\partial G}{\partial \xi}
\end{equation}
Therefore, this transport equation reveals that, if an input wave
energy pulse, $F_{I}(k,t=0)$, is initially centered at intermediate
$k$ values, say $k=k_{0}$, it is then carried \emph{conservatively}
at constant speed $\alpha\equiv V_{A}'$, on a logarithmic scale,
until it reaches the dissipative range $k=k_{d}$. \emph{Only there}
it experiences fast damping, and finally converts its energy into
heat. This spectral viewpoint also contains the time scale involved
before dissipation occurs : it is the travel time in the energy
conserving range in k-space, $\tau=(\ln k_{d}-\ln
k_{0})/V_{A}'=(2/V_{A}')\ln (V_{A}'/\eta k_{0}^{2})\propto \ln S$,
which is $\tau_{strain}$.

Before concluding, few comments are due. First, let us note that the
\emph{magnitude} of the energy flux $\epsilon$ depends on the amplitude of the
drive, but not on its frequency, see e.g. Velli (2003).
However, the location of the energy source in $k$-space depends on
the range of driving frequencies. Recall that since the dynamical
equations are linear, consideration of a broad frequency drive does
not alter the shape of the spectrum in the gap between the largest
input wavenumber and the dissipative one. The very existence of such
conservative range with neither source nor dissipation is suggested
by the following reasoning. Since the wave-number associated with
the dissipative scale is very large,
$k_{d}=(V_{A}'/\eta)^{1/2}\gg1$, the condition that the latter is
also much larger than the largest input wave-number $k_{\|
max}=\omega/V_{A}$, i.e. $k_{\| max}\ll k_{d}$, is equivalent to
$\omega\tau_{A}\ll L(V_{A}'/\eta)^{1/2}\sim S^{1/2}$. This condition can
easily hold for also for waves with $\omega\tau_{A}\gg1$ or
equivalently, $k_\|\gg L^{-1}$, for which the WKB approximation
holds. In their original work, Similon and Sudan (1989), were able
to relax the condition $k_{\|}\gg L^{-1}$, making it possible to
study the behavior of the lower frequency
fluctuations and "non-WKB" effects. Their tactic
is entirely analogous to the reduced magnetohydrodynamics (RMHD)
approximation, although they refer to the so-called "ballooning
representation" in fusion studies. By doing so, they only restrict
the eikonal representation to the perpendicular direction. Low
frequency motion at the boundary of a unidirectional magnetic field
is well known to produce turbulence and energy cascade leading to a
power-law energy spectrum [see for instance two recent works,
Rapazzo et al (2007), Dmitruk et al. (2003) and references therein]. For realistic forcing,
i.e. small amplitudes of the Alfv\'en waves that are launched into the
corona, $\delta u\sim 10^{-3} V_{A}$ (Acton et al. 1981), numerically obtained
spectra are generally found to be quite steep. The reason resides in the weakness
of the nonlinear interactions. However,
since the background large scale field is taken as a constant, only the
nonlinearity can there provide the necessary transfer of
energy to small dissipative scales. The linear mixing process with its associated energy flux
is likely to alter energy transfer
rates and hence spectra in inhomogeneous magnetic backgrounds.
This is suggested by the numerical studies by Malara et al. (1992), Einaudi et al. (1996),  who 
showed that scattering of Alfv\'en waves with the background inhomogeneity, for a population of 
waves initially propagating say, upward,
can act as a source for downward-propagating waves, and hence as a trigger for the occurrence 
of non-linear interactions.   
We did not consider such effect in the present work and we 
are not aware of any MHD turbulence theory that accounts for
how the linear and non-linear mechanisms of energy flux combine
together. We find this subject an interesting area for further research.

We finally mention that the mixing effect considered
in the present article, is the root of a proposed geometrical interpretation
for the consequence of non-linear
interactions among counter-propagating wave packets. It is clearly
illustrated by Fig.1 in Maron and Goldreich (2001). Indeed, take a
sample of originally straight vertical field lines which are
perturbed by, say, downward-propagating waves. As a consequence, the
field lines wander around each other. Therefore, if a
localized wave packet is launched and propagates upward in this
braided field, it will be distorted [also compare with Fig.2 in
Similon and Sudan (1989) or the figure in Rechester and Rosenbluth
(1977)]. After several of these collisions, dissipation takes over and the wave
packet mixes in its environment.

\section{Conclusions}
In this work, we consider the wave energy flux to small scales which
results from the distortion of Alfv\'en waves propagating in an
inhomogeneous mean magnetic field $\mathbf{B}_{0}$.
This problem is related to turbulence because the existence of a mean field may provide
separation between linear and non-linear time-scales. In an inhomogeneous magnetic field,
Alfv\'en wave packets disperse and are
stretched by an Alfv\'enic flow
$\mathbf{V}_{A}=\mathbf{B}_{0}/(\mu_{0}\rho)^{1/2}$, whose
streamlines are also the magnetics field lines. In a plasma with
uniform density $\rho$, this flow is incompressible, i.e.
$\nabla.\mathbf{V}_{A}=0$, and its stretching properties are thus
only related to magnetic field lines dispersion. While the wave
packets transport their energy at the group velocity
$\mathbf{V}_{A}$, and hence while they disperse, in physical space,
 as the field lines
do, they also experience continuous shearing and straining, i.e. they tend to mix.
This gives rise to a cascade of the wave energy in $k$-space. As is well
known for mixing, the efficiency of stretching relies on the local
exponential separation of nearby trajectories (Ottino 1988). Since 
for a continuous external wave energy drive, the cascade of this energy
ultimately leads to a balance between source and heating in the
dissipative range, a stationary wave energy spectrum can build up
even in the absence of significant nonlinear interactions among the waves.
A phenomenon which is absent in a constant mean magnetic field.
Assuming the existence of a conservative range, we
determined the wave-energy spectrum for the special case of wave
packets propagating in a "fully" chaotic magnetic field. It means
that the distribution of "Lyapunov exponents", representing the rate
of local exponentiation of nearby field lines, and hence the degree
of stretching of wave packets, is strongly peaked at an average
value $\alpha$. This situation, which was first considered by
Similon and Sudan (1989), for dissipation of Alfv\'en waves 
corresponds to chaotic mixing by flows with uniformly 
distributed rate of strain (Batchelor 1959, Antonsen et al. 1996).

\begin{acknowledgements}
This work was supported by the Science and Technology Facilities
Council (STFC)
\end{acknowledgements}

\end{document}